\documentclass[
preprint,
 amsmath,amssymb,
 prl,
 aps,
]{revtex4-2}
\usepackage{xcolor}
\usepackage{graphicx}
\usepackage{dcolumn}
\usepackage{bm}

\begin{document}
%\preprint{APS}

\title{Impact of Electric Spatially Discordant Alternans on Cardiac Magnetic Field}

\author{Martina Nicoletti\textsuperscript{1,2*}, Anna Crispino\textsuperscript{1*}, Alessandro Loppini\textsuperscript{3,4}$^\dagger$, Alessio Gizzi\textsuperscript{1}, Letizia Chiodo\textsuperscript{1,4}, Christian Cherubini\textsuperscript{5,6}, Simonetta Filippi\textsuperscript{1,4}}

\affiliation{\textsuperscript{1}Department of Engineering, Universit\`a Campus Bio-Medico di Roma, Italy
}%

\affiliation{\textsuperscript{2}Center for Life Nano and Neuro Science (CLN2S@Sapienza), Italian Institute of Technology, Rome, Italy
}%

\affiliation{\textsuperscript{3}Department of Medicine and Surgery, Università Campus Bio-Medico di Roma, Italy
}%
\affiliation{\textsuperscript{4}Istituto Nazionale di Ottica del Consiglio Nazionale delle Ricerche (CNR-INO), Florence, Italy
}%
\affiliation{\textsuperscript{5}Department of Science and Technology for Sustainable Development and One Health, Università Campus Bio-Medico di Roma, Italy}

\affiliation{\textsuperscript{6}ICRANet - International Center for Relativistic Astrophysics Network, Piazza della Repubblica, 10, Pescara, 65122, Italy}

\affiliation{*These authors contributed equally to this work}

\affiliation{\textsuperscript{$\dagger$}Corresponding author: a.loppini@unicampus.it}

\begin{abstract}
Spatially discordant alternans (SDA) play a crucial role in cardiac arrhythmogenesis by creating steep repolarization gradients facilitating conduction block and reentry. While traditionally studied using electrical indicators, this work provides a novel perspective by characterizing SDA through their magnetic field signatures. Using a one-dimensional cardiac fiber model, we demonstrate that magnetic field measurements effectively detect SDA and temperature dependent changes in cardiac action potentials, offering a non-invasive alternative to conventional electrophysiological metrics. Our results reveal that the spatial organization of SDA is mirrored in the magnetic field distribution, with SDA nodes clearly identifiable via spatial mapping. Notably, magnetic restitution curves exhibit a distinct pattern from APD-based indicators, closely following the dynamics of the action potential upstroke. These findings establish the cardiac magnetic field as a powerful diagnostic tool for detecting SDA, opening new avenues for biomagnetic monitoring of arrhythmic risk.
\end{abstract}

\maketitle

%\tableofcontents

\section{Introduction}\label{sec1}
Cardiac arrhythmias are among the leading causes of sudden cardiac death, significantly contributing to global mortality~\cite{DiCesare2024, Zheng2001, Kim2011}. Despite advancements in diagnosis and treatment, further investigation is needed, particularly to understand cardiac alternans, which serve as precursors to arrhythmias~\cite{Gizzi2013, Ito_1991, Courtemanche_1993, Fenton_2000, Hall_1997, Bub_2005}. Spatiotemporal alternans dynamics have been traditionally studied as beat-to-beat variations in action potential duration (APD) and classified as spatially concordant (SCA), when APD alternates in-phase across the tissue, or spatially discordant (SDA), when different regions alternate out-of-phase. The latter creates nodes where APD remains unchanged~\cite{Hall_1997, Uzelac2017SimultaneousTransients, Crispino2024cross, de2008spatially}. SDA are of particular interest due to their strong association with arrhythmogenesis~\cite{weiss2006pulsus, qu2023cardiac}.
%influenced by factors such as ion dynamics, tissue anisotropy and heterogeneity, and external conditions like thermal variations~\cite{kiyosue1993ionic, crozier1926distribution, FEDOROV20081587, Filippitemperature, Gizzi2013, correlation_temperature_2021, Crispino2024cross, crispino2024magnetic}.
Several factors contribute to SDA onset, including ischemic damage, intrinsic tissue heterogeneities, and emergent patterns in cells and tissues~\cite{Watanabe2001MechanismsAlternans, pastore1999mechanism, pastore2006importance, hayashi2007dynamic, watanabe2001mechanisms, qu2000mechanisms, Gizzi2013}. Among external modulators, temperature plays a crucial role, as lower temperatures prolong APD, increasing dispersion of repolarization and arrhythmic susceptibility~\cite{kiyosue1993ionic, crozier1926distribution, FEDOROV20081587, Filippitemperature, Gizzi2017, correlation_temperature_2021, Crispino2024cross}. While hypothermia is widely used in clinical settings to reduce cellular damage, it also presents pro-arrhythmic risks~\cite{MOORE2011, Rippe2018}.

Beyond traditional electrical measurements, the magnetic field generated by cardiac electrical activity provides valuable complementary insights~\cite{ROTH1986, ROTH1988, plonsey_nature_1982, Holtzer2004, roth2023biomagnetism}. However, the main challenge in cardiac magnetic field investigations lies in the extremely low intensity of these fields, typically in the pT–nT range~\cite{roth2024magnetocardiogram,baudenbacher2002high,roth2023biomagnetism,nakayama2015real,webb2021detection}, making them challenging to detect with conventional methods. Recent advances in quantum magnetometry, particularly Superconducting Quantum Interference Devices (SQUIDs)\cite{baudenbacher2002high, tanaka2003measurement, nakayama2015real} and Nitrogen-Vacancy (NV) centers in diamonds~\cite{barry2016optical, barry2020sensitivity, webb2021detection, arai2022millimetre}, have significantly enhanced the sensitivity of biomagnetic measurements, enabling the detection of ultra-weak signals even at the cellular level. This opens new perspectives for investigating the molecular mechanisms underlying arrhythmias, particularly in conditions like Catecholaminergic Polymorphic Ventricular Tachycardia (CPVT) and other rare genetic disorders.

Mathematical models have significantly contributed to expand our understanding of cardiac dynamics in health and disease, as they can offer deep insights on the biophysical mechanisms at the basis of cardiac arrhythmias onset and development~\cite{Gizzi2017, Filippitemperature,fenton1998vortex,fenton2013role}. In this work we aim to expand the domain of cardiac model application. Mathematical models of cardiac magnetic field~\cite{ROTH1988,crispino2024magnetic} highlighted its potential investigation demonstrating that magnetic field produced by cardiac activity could convey physical information that is not included in the sole electric field~\cite{ROTH1988}.

%In the present work, we analyze the initiation and progression of APD alternans in a one-dimensional homogeneous cable under varying thermal conditions, incorporating a novel magnetic field-based perspective~\cite{crispino2024magnetic}. 
In the present work, we analyze the initiation and progression of cardiac alternans in a one-dimensional homogeneous cable under varying thermal conditions, based on a thermo-electro-magnetic modeling framework recently developed~\cite{crispino2024magnetic}. In particular, we here apply this approach to the investigation of alternans and development of SDA regimes. Using standard pacing-down restitution protocols, we derive the magnetic equivalent of restitution curves and introduce a novel approach to characterizing SDA alternans based on magnetic field dynamics. This work provides a methodological framework for biomagnetic analysis of cardiac alternans, which can be extended to more complex spatial and temporal electrical patterns in the heart, ultimately contributing to a deeper understanding of arrhythmogenesis.

\section{Methods}\label{sec2}
%In this section we describe how the magnetic field produced by cardiac activity has been computed. First of all we describe the model of ca

\subsection{Cardiac Electrophysiology Modeling} \label{sec2_1}
We model cardiac action potential using a four-variable phenomenological model~\cite{Gizzi2013,crispino2024magnetic}, which is suitable for describing temperature-related effects on cardiac dynamics~\cite{fenton2013role} as well as for simulating full pacing down restitution protocols.
The model presents a different parameters set with respect to~\cite{crispino2024magnetic}, opportunely fine-tuned to reproduce SDA in one-dimensional cable geometries.
The membrane potential is described with the reaction-diffusion partial differential equation:
\begin{equation}
\frac{\partial u}{\partial t}= D\nabla^2u-(J_{fi}+J_{si}+J_{so}) 
\label{eq1}
\end{equation}
where, $u$ is the nodimensional membrane potential, $D$ is the diffusion coefficient, while $J_{fi}$, $J_{so}$, and $J_{si}$ represent the fast-inward, slow-outward, and the slow-inward current densities, respectively. The three currents are voltage ($u$) and temperature ($T$) dependent. Moreover, they are also regulated by three gate variables $v$, $w$, and $s$ through the following equations:
\begin{eqnarray}
\label{eq2}
J_{fi}&=&\eta_{fi}(T)\left[H(u-\theta_v)(u-\theta_v)(u-u_u)\frac{v}{\tau_{fi}}\right]\\
\label{eq3}
J_{si}&=&\eta_{si}(T)\left[H(u-\theta_w)\frac{ws}{\tau_{si}}\right]\\
\label{eq4}
J_{so}&=&\eta_{so}(T)\left[ (1-H(u-\theta_w))\frac{(u_o-u)}{\tau_{o}}+\frac{H(u-\theta_w)}{\tau_{so}}\right]
\end{eqnarray}
where $H$ indicates the Heaviside step function. The gating variables are obtained by solving the following coupled system of ordinary differential equations:
\begin{eqnarray}
\label{eq5} 
\frac{\partial v}{\partial t}&=&\Phi_v(T)\left[(1-H(u-\theta_v)) \frac{v_\infty -v}{\tau_v^-}-\frac{H(u-\theta_v)}{\tau_v^+} \right]\\
\label{eq6}
\frac{\partial w}{\partial t}&=&\Phi_w(T)\left[(1-H(u-\theta_w)) \frac{w_\infty -w}{\tau_w^-}-\frac{H(u-\theta_w)}{\tau_w^+} \right]\\
\label{eq7}
\frac{\partial s}{\partial t}&=& \Phi_s(T) \left[ \frac{(1+\tanh(k_s(u-u_s)))/2-s}{\tau_s} \right]
\end{eqnarray}
where, $w_\infty $ and $v_\infty$ are defined as:
\begin{eqnarray}
\label{eq8}
v_\infty&=&H(\theta_v^- -u)\\
\label{eq9}
w_\infty&=&(1-H(u-\theta_o)) \left( 1-\frac{u}{\tau_{w_\infty}}\right)+H(u-\theta_o)w_\infty^\star
\end{eqnarray}
The time constants ($\tau_{fi}$, $\tau_{si}$, $\tau_{so}$, $\tau_{o}$, $\tau_v^+$, $\tau_v^-$, $\tau_w^+$, $\tau_w^-$, $\tau_s$) in Eqs.~\ref{eq2}-\ref{eq7} are temperature and voltage dependent as follows: 
\begin{eqnarray}
\label{eq10}
\tau_v^-&=&(1-H(u-\theta_v^-))\tau_{v1}^{-}+-H(u-\theta_v^-)\tau_{v2}^{-}\\
\label{eq11}
\tau_w^+&=& \tau_{w1}^+ +(\tau_{w2}^+ - \tau_{w1}^+) \frac{\tanh(k_w^+(u-u_w^+))+1}{2}\\
\label{eq12}
\tau_w^-&=& \tau_{w1}^- +(\tau_{w2}^- - \tau_{w1}^-) \frac{\tanh(k_w^-(u-u_w^-))+1}{2}\\
\label{eq13}
\tau_{so}^-&=& \tau_{so1} +(\tau_{so2} - \tau_{so1}) \frac{\tanh(k_{so}(u-u_{so}))+1}{2}\\
\label{eq14}
\tau_s&=&(1-H(u-\theta_w))\tau_{s1}+H(u-\theta_w)\tau_{s2}\\
\label{eq15}
\tau_o&=&(1-H(u-\theta_o))\tau_{o1}+H(u-\theta_o)\tau_{o2}
\end{eqnarray}

Temperature effects on the membrane potential are modelled by modulating the current densities and the three gating variables $v$, $w$, $s$ with a temperature-dependent Moore ($\eta(T)$) and Arrhenius ($\Phi(T)$) factors, defined as follows~\cite{fenton2013role}: 
\begin{eqnarray}
\label{eqs16}
\Phi(T)&=&Q_{10}^{(T-T_A)/10}\\
\label{eqs17}
\eta(T)&=&A(1+B(T-T_A))
\end{eqnarray}

Model parameters are listed in Table~\ref{tab:table1}. We implemented the model in a straight cylinder ($L=3$~cm, $r=40~\mu$m, $dz=100~\mu$m) resembling a fibre of cardiac tissue.

\begin{table}[ht]
	
	\caption{Model parameters for endocardial action potential minimal-model formulation for a selected thermal range (29$\div$40$^\circ$C). Units are given in ms, cm, mV, mS, $\mu$F, g, $^\circ$C. We used as initial conditions $u=$, $v=1$,$w=0$, $s=1$ \cite{bueno2008minimal}.}
	\centering
	\begin{tabular}{|c|c||c|c||c|c|}
		\hline
		Parameter & Value & Parameter & Value & Parameter & Value \\ \hline
		$\theta_v$ & 0.3 & $\tau_{v^-_1}$ & 55 & $\tau_{w_1^+}$ & 175 \\ 
		$\theta_w$ & 0.13 & $\tau_{v^-_2}$ & 10 & $\tau_{w_2^+}$ & 230 \\ 
		$\theta_v^-$ & 0.2 & $\tau_v^+$ & 1.4506 & $\tau_{so_1}$ & 40 \\ 
		$\theta_o$ & 0.006 & $\tau_{w^-_1}$ & 6 & $\tau_{so_2}$ & 1.2 \\ 
		$\tau_{fi}$ & 0.10 & $\tau_{o1}$ & 470 & $k_{so}$ & 2 \\ 
		$\tau_{o2}$ & 6 & $u_w^+$ & 0.0005 & $u_{so}$ & 0.65 \\ 
		$k_w^-$ & 200 & $\tau_{s1}$ & 2.7342 & $\tau_{s2}$ & 2 \\ 
		$k_w^+$ & 8 & $k_{s}$ & 2.0994 & $u_{s}$ & 0.9087 \\ 
		$u_w^-$ & 0.00615 & $\tau_{si}$ & 2.9013 & $\tau_{w_\infty}$ & 0.0273 \\ 
		$w_\infty^\star$ & 0.78 & $Q_{10,v}$& 1.5 & $Q_{10,w}$ & 2.45 \\ 
		$ Q_{10,s}$ & 1.5 & $A_{fi}$& 1 & $B_{fi}$ & 0.065\\ 
		$A_{so}$ & 1 & $B_{so}$& 0.008 & $A_{si}$ & 1\\ 
		$B_{si}$ & 0.008 & & & &\\
		\hline
	\end{tabular}
	\label{tab:table1}
\end{table}

This simple--yet realistic--geometry is selected to keep the complexity of magnetic field calculation as low as possible while preserving a link with physiological case studies~\cite{crispino2024magnetic,bueno2015basis}. We study the system behaviour during a full pacing-down restitution protocol specifically designed to study spatially discordant alternans. During the protocol, the wire is stimulated with current pulses while decreasing the period, i.e., the pacing cycle length (PCL).
The protocol starts with sinus rhythm gradually decreasing until conduction block arises. Simulations have been performed at different temperature between 29$^\circ$C and 40$^\circ$C.

\subsection{Current density estimation}\label{sec2_2}
We compute the current originating the magnetic field from the simulated electrical activity applying the classical electromagnetism laws: 
\begin{equation}
\label{eqs18}
J(t,z')=-\sigma \nabla V_m(t,z'), \rightarrow J(t,z')=-\sigma \frac{\partial V_m}{\partial z'}
\end{equation}
where, $V_m(t,z')$ represents the transmembrane potential obtained from its dimensionless equivalent $u$ by using the map $V_m(t,z)= 83.3u(t,z')-V_{rest}$, $V_{rest}=-84$~mV is the typical resting potential of cardiac cells, and $83.3$ is a scaling factor used to adjust the amplitude of the action potential. Tissue conductivity, $\sigma=D S_0 C_m$, is estimated as in~\cite{fenton1998vortex}, assuming that our fiber is homogeneous and isotropic
%
%\begin{equation}
%    \label{eqs19}
%    \sigma=D S_0 C_m,
%\end{equation}
%
with $C_m=1~\mu$F/cm\textsuperscript{2} the membrane capacitance per unit area, $D=0.005$~cm/ms the diffusion coefficient, and $S_0$ the surface to volume ratio estimated through a standard dimension of a myocyte in cardiac fibers, i.e., $R'=40~\mu$m, $L=100~\mu$m. 

%\textcolor{red}{vedere se tenere qui o spostare solo sotto}We consider the current density uniformly distributed on the circular section of the fiber ($R'=40\mu$~m, $S'=\pi R'^2$), so that $\vec J\equiv(0,0,J_z)$. By virtue of this assumption we can compute the total current as follows:

%\begin{equation}
%\label{eqs20}
%   I(t,z')=\int_{S'}\vec J(t,\vec r')\cdot d\vec S'= 
%J(t,z')\pi R'^2
%\end{equation}

As usual in the Literature, in the above equations and in the following, primed quantities are referred to positions within the current source, i.e., the cylinder on which we simulate the electrical activity. 

\subsection{Magnetic Field Modeling}\label{sec2_3}

The current density $J(t,z')$ obtained as described in section~\ref{sec2_2} is used to compute the magnetic field due to the elctrical activity in the wire. Given the slow variations of the current compared to signals travelling at the speed of light, we can safely neglect electromagnetic wave phenomena and use standard magneto-static tools to compute the magnetic field. Taking into consideration the particular geometry of the system, i.e., a straight cylinder in the $z$-direction with $z'\in[z'_{min},z'_{max}]=[0, 3]$~cm, we compute the magnetic field generated by the source applying the Biot-Savart's law assuming the current density uniformly distributed on a circular cross section of the wire with a radius $R=40~\mu$m (approximately the radius of cardiac myocyte). Due to the assumption we made on the current density, the magnetic field produced by the propagation of cardiac action potential along the wire only preserves $x$ and $y$ components, and can be obtained computing the following 1D integrals:
\begin{equation}
\vec B
=
%    &-&
-
\frac{\mu_0}{4\pi}\int_{z'_{\min}}^{z'_{\max}}\frac{I(t,z')y\,dz'}{[x^2+y^2+(z-z')^2]^{\frac32}} \,\hat i
%    \nonumber\\ 
+ 
\frac{\mu_0}{4\pi}\int_{z'_{\min}}^{z'_{\max}}\frac{I(t,z')x\,dz'}{[x^2+y^2+(z-z')^2]^{\frac32}} \,\hat j
\label{eqs24}
\end{equation}
with $\hat{i}$ and $\hat{j}$ indicating the $x$ and $y$ versors, respectively. For further details on the calculation of the magnetic field we refer the reader to~\cite{crispino2024magnetic}.

Equation~\eqref{eqs24} is evaluated in a cylindrical domain in the vacuum. In this regard, it is worth mentioning that the tissue is immersed in saline conductive solutions, in general. The electrical activity inside the tissue could induce currents in the extracellular medium which could partially cancel the magnetic field produced by intracellular currents~\cite{SWINNEY1980719}. In the present work we compute the magnetic field within a distance of $50~\mu$m from the source. Accordingly, the effects of extracellular currents can be safely neglected because the action potential wavelength ($\lambda_{AP}\simeq 10\div20$~cm) is significantly larger than the distance at which the field is evaluated~\cite{SWINNEY1980719, barry2016optical}.

\subsection{Cardiac alternans indicators}
We analyze the pro-arrhythmic phenomenon of cardiac alternans both from the electric and magnetic point of view, comparing the traditional indices-- as the action potential duration (APD)-- with novel magnetic ones. % To investigate them from the electrical point of view we use a standard indicator: the action potential duration (APD\textsubscript{80}). 

For each PCL, we compute the APD\textsubscript{80} of the last two beats thresholding the simulated AP at 80\% of repolarization. 
%In this way, we obtain APD-PCL restitution curves which are suitable for highlighting the presence of APD alternans.   
The magnetic activity associated with AP propagation is investigated by computing the maximal value of the magnetic field norm at a fixed distance from the wire ($50~\mu$m). In this regard, it is essential to note that the norm of the magnetic field is not the direct equivalent of the APD; rather, it is related to the time derivative of the action potential. Indeed, for electrical signals with a constant conduction velocity ($v_c$), the following equality between the spatial and temporal derivatives holds:
\begin{equation}
\frac{\partial V_m(t,z')}{\partial t}=-v_c \frac{\partial V_m(t,z')}{\partial z'}
\label{eqs25}
\end{equation}

In particular, the maximal value of the magnetic field norm is directly proportional to the maximal value of the time derivative of the action potential, which occurs around the upstroke, as detailed in~\cite{crispino2024magnetic} and discussed in the results. %Similar to electrical restitution curves, the maximal norm of the magnetic field has been computed for each PCL.  
%The aforementioned electrical and magnetic indicators have been employed to characterize AP alternans, with a particular focus on SDA.  

Cardiac alternans have been analyzed using restitution curves, which depict the local values of APD\textsubscript{80} or $|\mathbf{B}|_{\max}$ as a function of PCL. Unlike our previous work~\cite{crispino2024magnetic}, we chose to plot pointwise restitution curves to preserve the intrinsic variability of the data, as spatial averaging, particularly in the presence of SDA along the wire, could artificially attenuate the true extent of alternans dynamics.  

Spatially discordant alternans were further investigated by analyzing the spatial distribution of APD\textsubscript{80} and $|\mathbf{B}|_{\max}$ along the wire for the last two beats of each PCL (\( n-1 \) and \( n \)). APD\textsubscript{80}-spatial maps were constructed by computing APD\textsubscript{80} at each position along the wire. Similarly, $|\mathbf{B}|_{\max}$-spatial maps were obtained by averaging the maximal value of the magnetic field norm for each selected beat over the points in the $xy$-plane located at a distance of $50 \pm 1~\mu$m from the current source.

%For the sake of clarity, we report the APD\textsubscript{80} and $|\mathbf{B}|_{max}$ - space curves for three selected PCL representative of sinusoidal rhythm, alternans onset, and alternans regimes. 

%. We also analyse spatially discordant alternans from the point of view of the magnetic field. For this purpose, for each $z$ coordinate along the wire, we compute the average value of the temporal maximum of the magnetic field in the points of the $xy$-

\subsection{Numerical Methods}
The minimal model for the simulation of the cardiac electrical activity described in section~\ref{sec2_1} has been implemented in MATLAB R2022b (MathWorks, Natick, Massachusetts) using the finite-difference method to approximate both the temporal and spatial derivatives and solved with the Euler scheme. The final geometry consists in a 3~cm long wire discretized with a mesh containing 300 nodes. The number of nodes has been optimized with a convergence analysis to ensure the stability of the conduction velocity. The temporal discretization has been selected to ensure an accurate description of the AP upstroke, and therefore of the temporal characteristics of the magnetic field~\cite{crispino2024magnetic}. 
The magnetic field, Eq.~\eqref{eqs24}, is computed on the cylindrical domain ($R=300~\mu$m, $L=3$~cm) using the trapezoidal approximation method and discretized with a $10~\mu$m spatial grid. 

\section{Results}
%In this section, we present the main findings of this study, summarized as follows:
%\begin{itemize}
%\item[-] a comparison between action potential, current density, and the magnetic field;
%\item[-] a comparative analysis of AP- and $|\mathbf{B}|_{\max}$-restitution curves;
%\item[-] a characterization of spatially discordant alternans using both conventional electrical and magnetic indicators.
%\end{itemize}

In this section we present the main findings of the work. 
In Sec.~\ref{res_1} we analyze the model here refined to reproduce SDA in terms of its ability in reproducing the relation between cardiac action potential and the corresponding magnetic field. Specifically, we describe the spatial and temporal distribution of membrane potential, current density and magnetic field.
%In Sec.~\ref{res_1} we describe the relation between cardiac action potential and the corresponding magnetic field, describing in particular the spatial and temporal distribution of membrane potential, current density and magnetic field. 
In Sec.~\ref{res_2} we explore the influence of the thermal state on the action potential amplitude and the magnetic field presenting both the classical APD-PCL restitution curves, and the novel $|\mathbf{B}|$-PCL curves. Finally, in Sec.~\ref{res_3} we demonstrate the magnetic field analysis as a reliable tool for the investigation of spatially discordant alternans.

\subsection{Electric and magnetic description of cardiac APs} \label{res_1}
\begin{figure}

	\includegraphics[width=\linewidth]{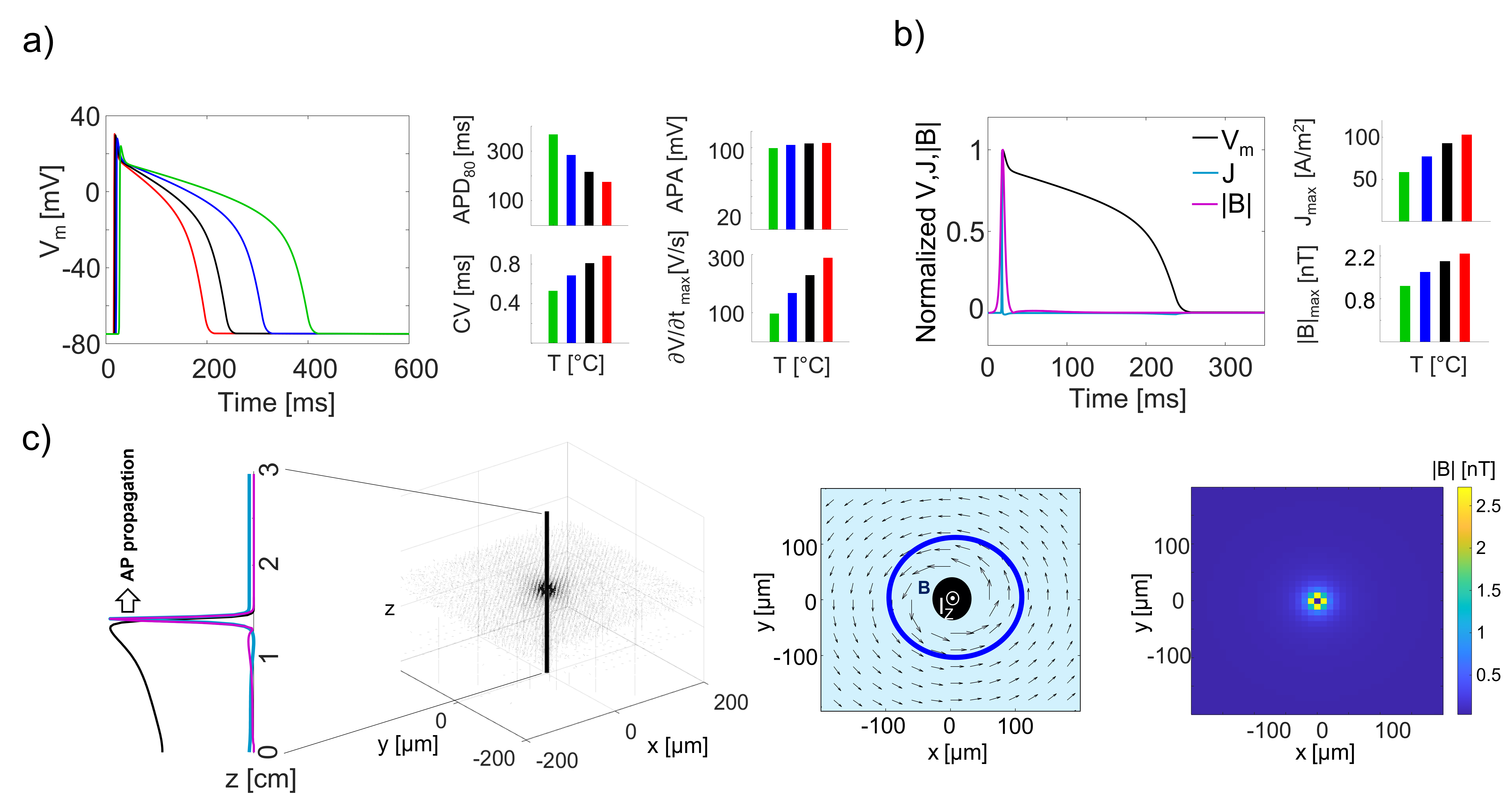}
	\caption{\footnotesize{\textbf{Simulation configuration.} \textbf{a) Temperature-induced changes in the cardiac action potential}. The panel reports the time course of the cardiac action potential for a fixed location in the centre of the wire at the three selected temperatures: 29$^{\circ}$C in green, 33$^{\circ}$C in black, 37$^{\circ}$C in blue, and 40$^{\circ}$C in red. The bar plots on the right report the action potential duration, the conduction velocity, the action potential amplitude (APA), and the maximum of the AP time derivative $\partial V/\partial t_{\max}$ for the action potentials shown on the left. \textbf{b) Relation between action potential, current density, and magnetic field.} In the panel we show the normalized temporal behavior of the action potential (in black), current density (in violet), and magnetic field (in cyan) at 37$^{\circ}$C for a fixed location in the middle of the wire. The current density is computed from the spatial derivative of the AP using Eq.~\eqref{eqs18} and the magnetic field is obtained from Eq.~\eqref{eqs24}. The reference action potential is already described for panel a). \textbf{c) Spatial configuration of the magnetic field}. In this panel we resume the main spatial features of the magnetic field produced by the cardiac APs. On the left we show the normalized spatial distribution along the wire of the membrane potential, current density and of the magnetic field at a fixed time frame. In the middle we show the vectorial representation of the magnetic field along the wire at a fixed time point. The insert on the right shows the 2D representation of the magnetic field in the $xy$-plane, with the blue ring indicating the distance from the wire at which we computed the norm of the magnetic field. Finally, on the right we report a colormap showing the spatial distribution of the magnetic field intensity in the $xy$-plane. All the plots in panel c) are obtained at 37$^{\circ}$C in the same simulation conditions of panel a). }}
	\label{fig:fig1}
\end{figure}

We investigate the temperature dependence of the electrophysiological properties of cardiac action potentials and their relationship with the magnetic field. To this end, we focus on the temporal and spatial dynamics of electrical and magnetic activity under normal sinus rhythm conditions (Fig.~\ref{fig:fig1}). Specifically, we stimulate the tissue with a pacing cycle length (PCL) of 800 ms delivering 15 stimuli and we analyze the last beat of the sequence.

As shown in Fig.~\ref{fig:fig1}-a), temperature has a significant influence on the shape of the action potential, particularly affecting its duration, quantified by APD\textsubscript{80}, and its upstroke, characterized by the maximum rate of voltage change ($\partial V/\partial t_{\max}$). In agreement with experimental data \cite{fenton2013role, Filippitemperature}, at physiological temperature (T=$37^\circ$C), the action potential duration is approximately 250 ms, with an amplitude of 100 mV and an upstroke velocity of 228 V/s. As the temperature decreases, the action potential duration progressively increases, nearly doubling at 29$^\circ$C, while the upstroke velocity slows down to 96 V/s. Conversely, under hyperthermic conditions (T=$40^\circ$C), APD decreases below 200 ms whereas the upstroke velocity increases. In contrast to APD and $\partial V/\partial t_{\max}$, the action potential amplitude (APA) remains largely unaffected by temperature variations. Notably, the thermal state of the tissue strongly impacts conduction velocity, with higher CVs at higher temperatures (Fig.~\ref{fig:fig1}-a)).

Among the classical parameters used to characterize the action potential, the upstroke velocity plays a critical role in explaining the relationship between cardiac electrophysiology and the magnetic field generated by electrical activity. For electrical signals propagating at a constant velocity, the magnetic field is directly linked to the temporal derivative of the action potential, as described by Eqs.~\eqref{eqs18}-\eqref{eqs25}. Specifically, the current density, \( J \), depends on the spatial derivative of the action potential, Eq.~\eqref{eqs18}, which in turn is related to its temporal derivative, as outlined in Eq.~\eqref{eqs25}. As illustrated in Fig.~\ref{fig:fig1}-b), the current density exhibits a sharp peak during the upstroke, where the temporal derivative of the action potential reaches its maximum. In contrast, during the repolarization phase, the current density rapidly decreases to zero, displaying a small negative peak toward the end of the action potential. As expected, the current density follows the same temperature dependence as $\partial V/\partial t_{\max}$, i.~e., increasing at higher temperatures.

According to these observations, the total current density responsible for generating the magnetic field increases with temperature increment. As a direct result, the magnetic field intensity also increases, reaching a peak value of approximately 1.8 nT at 37$^\circ$C at a distance of 50~$\mu$m from the source. Similar to the current density, the magnetic field reaches its maximum during the upstroke and subsequently declines to zero throughout the plateau and repolarization phases.

As the final step of this preliminary analysis, we examined the spatial distribution of the membrane potential, current density, and magnetic field (Fig.~\ref{fig:fig1}-c)). The current density and the corresponding magnetic field reach their maximum values on the wavefront (depolarization phase). Due to the intrinsic symmetry of the problem, the magnetic field is confined to a narrow region along the \( z \)-axis, where its peak is localized. As expected, the magnetic field streamlines form concentric circles centered in the wire and oriented according to the direction of the current. Moreover, in agreement with the laws of magnetostatics, the magnetic field intensity decreases with increasing distance from the wire (Fig.~\ref{fig:fig1}-c)).  

\subsection{Electric and magnetic alternans}\label{res_2}

After investigating the relation between the action potential and the magnetic field under normal rhythm conditions, we analyzed the electrical and magnetic activity of the tissue in the alternans regime. To characterize alternans dynamics, we report four key indicators: APD\textsubscript{80}, maximum upstroke velocity ($\partial V/\partial t_{max}$), conduction velocity (CV), and the peak value of the magnetic field norm.

\begin{figure}[!h]
	\centering
	\includegraphics[width=\linewidth]{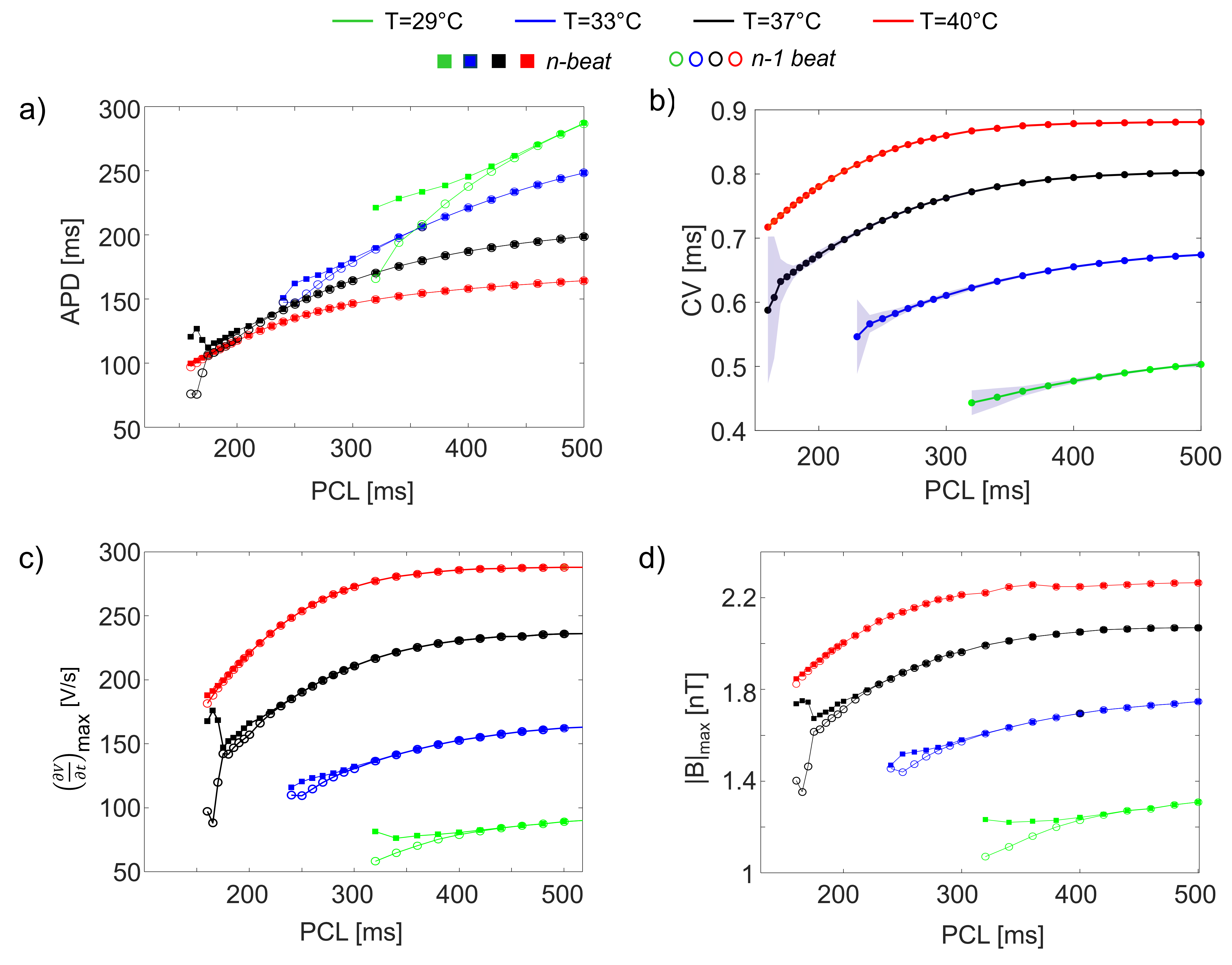}
	\caption{\footnotesize{\textbf{Action potential and magnetic field.} \textbf{(a) APD\textsubscript{80} restitution curves}. This panel presents classical restitution curves, illustrating the variations in action potential duration as a function of pacing cycle length (PCL) and tissue temperature. The curves are color-coded as reported in the legend. \textbf{(b) Conduction velocity restitution curves}. This panel displays the average conduction velocity (CV) computed for each temperature and PCL. The shaded regions represent the variation in CV between beats \(n\) and \(n-1\). The average CV is obtained by averaging the conduction velocities of the \(n\) and \(n-1\) beats. \textbf{(c) Upstroke velocity restitution curves}. This panel reports the maximum value of the time derivative of the action potential ($\partial V/ \partial t_{\max}$) for each PCL in the restitution protocol, evaluated for beats \(n\) and \(n-1\). \textbf{(d) $|\mathbf{B}|$\textsubscript{max} restitution curves}. The magnetic field norm restitution curves follow the same color scheme as panel (a) to indicate different temperatures.} %For further details on the methodology, we refer the reader to the \textit{Methods} section.
	} 
	\label{fig:fig2}
\end{figure}

Action potential alternans were investigated by stimulating the sample using a full pacing-down restitution protocol, as described in the Methods section. Consistent with previous studies~\cite{Crispino2024cross,fenton2013role}, lower temperatures promote the onset and progression of cardiac alternans (Fig.~\ref{fig:fig2}-a)). APD restitution curves indicate that at 29$^{\circ}$C, alternans emerge for PCLs below 340 ms, whereas at physiological temperature (37$^\circ$C), the onset of alternans is significantly shifted to shorter PCLs ($\text{PCL}_{\text{onset}}$ = 175~ms). At the highest temperature analyzed (40$^{\circ}$C), no alternans are detectable in the APD, even at low PCLs.

The other electrophysiological parameters considered, namely the maximum upstroke velocity ($\partial V/\partial t_{\max}$) and the conduction velocity, exhibit the same behavior as APD during the pacing-down restitution protocol, as shown in Fig.~\ref{fig:fig2}-b)-c). At this stage, we do not report the action potential amplitude as a function of PCL, since, as discussed, this parameter is not significantly affected by temperature variations.

We characterize the alternans regime from the perspective of the magnetic field by constructing the magnetic restitution curves shown in Fig.~\ref{fig:fig2}-d). These curves illustrate the relationship between the maximal value of the magnetic field at a distance of 50~$\mu$m from the source and the corresponding PCL. Our results suggest that AP alternans can be detected from the magnetic restitution curves, with the onset of alternans being more easily identifiable in these curves than in those of APD. Notably, the magnetic restitution curves appear in reverse order compared to the corresponding APD curves, while they closely follow the behavior of $\partial V/\partial t_{\max}$ (Fig.~\ref{fig:fig2}-b)-c)-d)), as expected from Eqs.~\eqref{eqs18}-\eqref{eqs24}.

\subsection{Electric and magnetic SDA} \label{res_3}
\begin{figure}[!h]
	\centering
	\includegraphics[width=\linewidth]{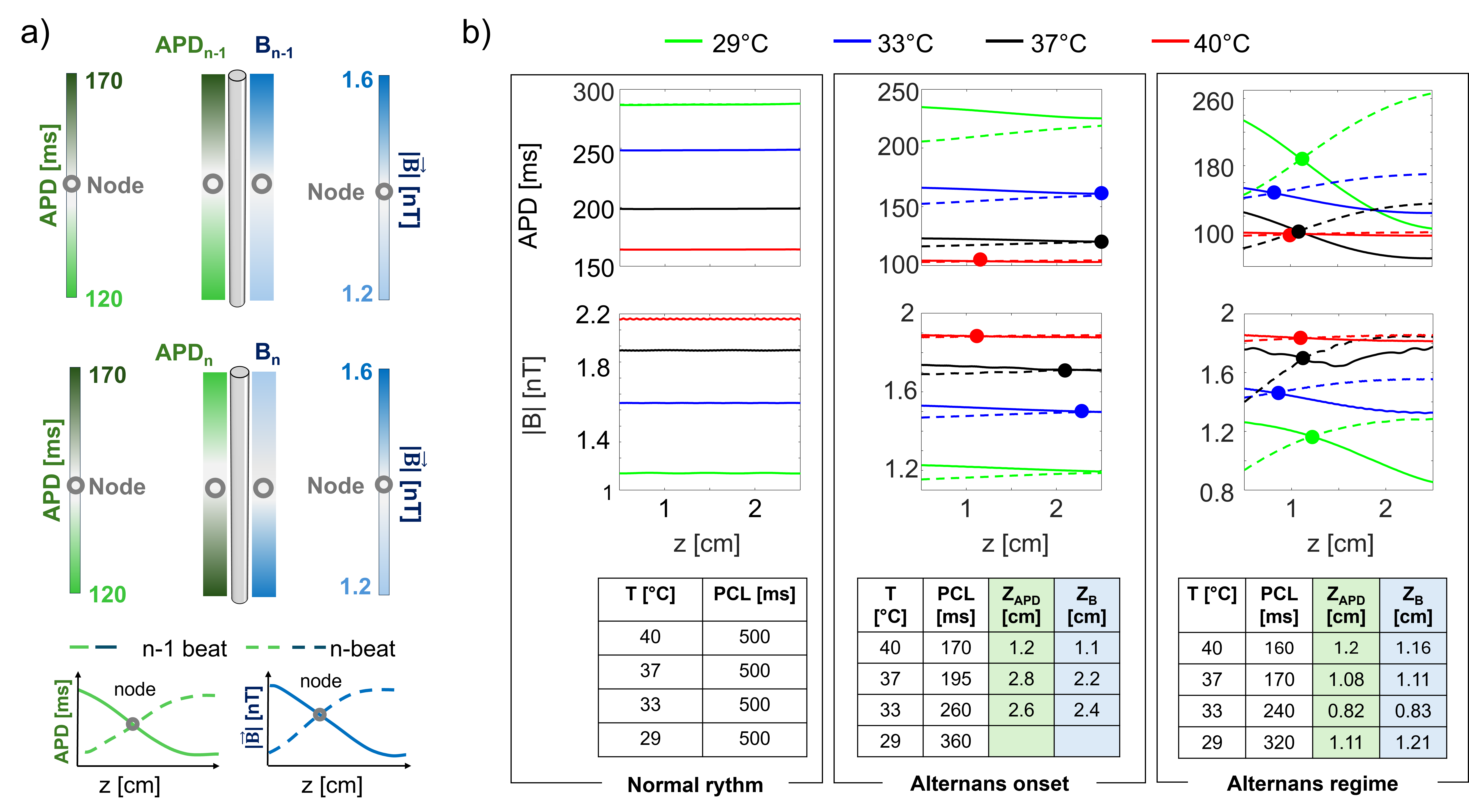}
	\caption{\footnotesize{\textbf{Spatially discordant alternans.} \textbf{(a) APD and $|\mathbf{B}|_{max}$ SDA}. Representation of the spatial alternans map for Action Potential Duration (APD) and the magnetic field norm \(|\mathbf{B}|_{\max}\) along the wire for two consecutive beats (\(n-1\) and \(n\)). \textbf{(b) Temperature dependence of SDA.} The spatial distributions of APD and \(|\mathbf{B}|_{\max}\) are illustrated for each temperature across three distinct electrophysiological regimes: sinus rhythm, alternans onset, and fully developed alternans. Pacing cycle lengths (PCLs) and the corresponding spatial locations of the node for both measured parameters (\(\mathrm{Z_{APD}}\) and \(\mathrm{Z_B}\)) are summarized in the tables at the bottom. Circles denote spatial nodes where the electrical and magnetic properties assume equal values between the selected beats. Empty cells denote the absence of the node.}}
	\label{fig:fig3}
\end{figure}

Spatially discordant alternans (SDA) are characterized by different regions of the tissue exhibiting out-of-phase action potentials, leading to dispersion of repolarization and enhanced tissue heterogeneity~\cite{Gizzi2013,crispino2024magnetic}.
As the final step of our work, we analyze the spatial behavior of alternans (Fig.~\ref{fig:fig3}) to investigate whether temperature-dependent SDA arise in a simple fiber geometry and whether they can be detected via magnetic field measurements. To this end, we compute the APD and the magnetic field norm along the length of the wire at three selected PCLs, representative of normal rhythm, alternans onset, and the fully developed alternans regime. Given the temperature-dependent variability in alternans onset, the PCLs corresponding to alternans onset and the alternans regime are specifically selected for each temperature, see Fig~\ref{fig:fig3}.

As expected, under normal rhythm conditions, the APD and the magnetic field norm assume the same value for beats \(n\) and \(n-1\) at each position along the wire. At alternans onset, SDA emerge at 33$^{\circ}$C, 37$^{\circ}$C and 29$^{\circ}$C, while for 40$^{\circ}$C, all regions of the domain do not show measurable alternans (Fig.~\ref{fig:fig3}-b)).  
%Spatially discordant alternans persist at 33$^{\circ}$C and 37$^{\circ}$C even in the fully developed alternans regime. 
Notably, SDA are also clearly detectable by analyzing the magnetic field norm as a function of space. The magnetic field curves closely mirror the APD ones, exhibiting spatially discordant alternans at 33$^{\circ}$C and 37$^{\circ}$C (Fig.~\ref{fig:fig3}-b)).

The space-dependent positions of the nodes in spatially discordant alternans, denoted as \(\mathrm{Z_{APD}}\) and \(\mathrm{Z_B}\), are summarized in Fig.~\ref{fig:fig3} for different temperatures and PCLs. At 37$^\circ$C and 33$^\circ$C, corresponding to PCLs of 170~ms and 240~ms, respectively, the node positions for APD and the magnetic field norm value show close agreement, with deviations of only a few millimeters (\( \mathrm{Z_{APD}} = 1.08 \, \mathrm{cm} \), \( \mathrm{Z_B}  = 1.11 \, \mathrm{cm} \) at 37$^\circ$C; \( \text{Z}_{\text{APD}} = 0.82 \, \mathrm{cm} \), \(\mathrm{Z_B}  = 0.83 \, \mathrm{cm} \) at 33$^\circ$C). This strong spatial correlation between electrical and magnetic indicators highlights the capability of magnetic field measurements to reliably detect SDA.
Similarly, at lower temperatures (29$^\circ$C) and higher PCL (320~ms), the nodes shift further along the wire (\( \text{Z}_{\text{APD}} = 1.11 \, \mathrm{cm} \), \( \mathrm{Z_B} = 1.21 \, \mathrm{cm} \)). This displacement indicates a temperature-dependent variation in the dynamics of SDA, as lower temperatures are associated with a broader spatial distribution of alternans. 
%Interestingly, at 40$^\circ$C, no nodes are observed, as SDA are absent at this temperature, further confirming the inhibitory effect of high temperature on the onset and persistence of SDA.

These results demonstrate the sensitivity of both electrical and magnetic indicators to spatial alternans patterns, emphasizing the benefit of magnetic field measurements for non-invasive detection of SDA in cardiac tissue.

\section{Discussion and Conclusion}
It is well known that variations in the thermal state of cardiac tissue influence action potential morphology and conduction velocity entailing an increased risk of arrhythmias at low temperatures. In this work, we explored thermo-electro-magnetic effects by analysing cardiac alternans from a magnetic point of view in the case of a one-dimensional domain representative of a cardiac fibre.

We started investigating the relation between the classical electrophysiological parameters, i.e., the action potential duration (APD), the action potential upstroke ($\partial V/\partial t_{max}$), and the action potential amplitude (APA), and the magnetic field produced by the electrophysiological activity. 
%In agreement with the literature, our simulations show action potentials with increased duration at low temperatures (Fig.\ref{fig:fig1}-a)). As the temperature increases, the action potential becomes shorter and the conduction velocity increases up to 0.8~m/s~(Fig.\ref{fig:fig1}-a)). Exploiting the laws of magnetostatics (Eq.~\ref{eqs18}-\ref{eqs24}), we computed the magnetic field generated by cardiac electrical activity (Fig.\ref{fig:fig1}-b), c)). 
The simulated magnetic field showed a marked thermal dependence, with intensity decreasing as the temperature decreases. 
%The norm of the magnetic field computed at a fixed distance from the source shows a temperature dependence, with increased intensities at high temperatures (Fig.\ref{fig:fig1}-b)). 
Such a trend is consistent with the temperature dependence of the upstroke velocity suggesting that the norm of the magnetic field might be used as a reliable indicator for investigating cardiac action potential features, in particular for a fine characterization of the upstroke phase. 

The integrative relation between the magnetic field and the action potential may amplify small alterations of the upstroke which might not be easily detectable with standard electrophysiological measurements, offering new interesting opportunities for gaining deeper insights on the molecular mechanisms of cardiac channelopathies involving mutations of the sodium channels, such as Brugada Syndrome, Isolated Progressive Cardiac Conduction Disease (PCCD), or Long-QT 3 syndrome~\cite{webster2013update}. Instead, the strict relation between the time derivative of the action potential and the magnetic field suggests that magnetic measurements could be less informative for the investigation of the plateau and repolarization phases of the action potential, because current density, and therefore the magnetic field, is almost null. 

Besides, we tested the norm of the magnetic field as a novel biomarker for detecting cardiac alternans. For this purpose, we analyzed the electric and magnetic behavior of the system during a full pacing-down restitution protocol to characterize the onset and development of cardiac alternans. Our results suggest that magnetic-based cardiac alternans can be detected at all temperatures. Overall, magnetic restitution curves display the same temperature-dependent behavior of the $\partial V/\partial t_{max}$ traces while they are inverted with respect to the classical APD restitution curves. Though differences between $n$ and $n-1$ beats are not significantly amplified in the magnetic field curves, the bifurcation point is more easily detectable than in the APD restitution curves, suggesting that magnetic field measurements could be a valuable alternative for the identification of arrhythmias precursors. 

Finally, we exploited the capability of the magnetic field to detect spatially discordant alternans. Also in this case, we demonstrated that a simple magnetic indicator, the norm of the field, is capable of detecting complex spatial dynamics that could result into fatal arrhythmias. 

%SDA have been suggested to be precursors of arrhythmias because they promote wave break and reentry by leading the formation of large voltage gradients in the myocardium~\cite{sato2014sda,pastore1999mechanism}. However, their molecular origin is still poorly investigated due to the complexity of the mechanisms involved. 

%In this context magnetic field measurements could offer new perspectives for the investigation of this phenomenon, especially in light of the fact that, besides the analysis performed on the norm of the magnetic field here performed, magnetic fields could convey physical information which is not included in the sole electric field. 

Before concluding, it is worth highlighting some critical aspects of the work. The first and the main limiting factor is the simplicity of the selected geometry, a 1D cable, which intrinsically restricts the variety of the spatiotemporal phenomena observed. The second limitation concerns the homogeneity of the sample, which does not reflect physiological variability and is argued to play a critical role in the onset and progression of cardiac alternans. Both these aspects will be considered for future extensions of the present study.

In conclusion, this work provides a first approach to analyzing cardiac action potential alternans using magnetic-field-based biomarkers, offering new perspectives for future investigation of complex spatiotemporal dynamics occurring in arrhythmogenic regimes.

\paragraph*{Funding \& Acknowledgments.}
This research has been funded by the European Commission-EU under the HORIZON Research and Innovation Action MUQUABIS GA n. 101070546, and by the European Union - NextGeneration EU, within PRIN 2022, PNRR M4C2, Project QUASAR 20225HYM8N [CUP C53D2300140 0008]. S.F. acknowledges the International Center for Relativistic Astrophysics Network - ICRANet. Authors wish to acknowledge the Italian National Group for Mathematical Physics, GNFM-INdAM.

\paragraph*{Author contribution statements.}
M.N. and A.C. contributed equally to this work.

\bibliography{biblio_new}% common bib file
\end{document}